\documentstyle[prl,twocolumn,aps]{revtex}

\begin{document}
\draft

\title{Phase diagram of depleted Heisenberg model for
${\rm CaV}_4{\rm O}_9$}

\author{Matthias Troyer, Hiroshi Kontani and Kazuo Ueda}

\address{
Institute for Solid State Physics, University of Tokyo, Roppongi 7-22-1,
Minatoku, Tokyo 106, Japan
}

\date{Received}
\maketitle

\begin{abstract}
We have numerically investigated the $1/5$-depleted Heisenberg square
lattice representing ${\rm CaV}_4{\rm O}_9$ using the Quantum Monte
Carlo loop algorithm. We have determined the phase diagram of the model
as a function of the ratio of the two different couplings: bonds within
a plaquette and dimer bonds between plaquettes. By calculating both the
spin gap and the staggered magnetization we determine the range of
stability of the long range ordered (LRO) phase. At isotropic coupling LRO
{\it survives} the depletion. But the close vicinity of the isotropic
point to the spin gap phase leads us to the conclusion that already a
small frustrating next nearest neighbor interaction can drive the system
into the quantum disordered phase and thus explain the spin gap behavior
of ${\rm CaV}_4{\rm O}_9$.
\end{abstract}

\pacs{PACS numbers: 75.10.Jm, 75.30.Kz, 75.40.Mg, 75.50.Ee, 75.40.Cx}


The stability of the long range ordered (LRO) ground state of the planar
Heisenberg model has been at the focus of investigations for a long
time. The recent discovery of a spin gap in ${\rm CaV}_4{\rm O}_9$
\cite{taniguchi} has given special importance to this question.
This compound can be described by a $1/5$-depleted planar
antiferromagnetic Heisenberg model\cite{ueda,katoh}. One of the
important questions regarding this material is whether the depletion of
the square lattice can account for the spin gap, or if additionally
frustration effects are important.

The role of lattice defects and depletion in destabilizing LRO has been
studied in a variety of contexts. One example are spin ladders which can
be obtained from the planar copper oxide materials, by breaking up the
planes into ladders of constant width \cite{ladderreview}.
Another way to destroy LRO is to deplete the
lattice. The bonds between spins are then weakened, similar to the
introduction of holes, and quantum fluctuations are enhanced, which
might destroy LRO. An example is the triangular Heisenberg
antiferromagnet, which exhibits LRO. Depeletion of $1/4$ of the spins,
leads to the Kagome-lattice, which is believed to have no LRO
\cite{zeng}.

The stability of LRO is also of relevance in the field of high
temperature superconductors. There the rapid destruction of LRO upon
hole doping and the possibility of realizing a doped resonating valence
bond (RVB) phase \cite{anderson}, exhibiting a finite gap in the spin
excitation spectrum (spin gap) are of great current interest. The study
of lattice defects, such as depletion or the formation of ladders can
give valuable insights \cite{stripes}.

The lattice structure of ${\rm CaV}_4{\rm O}_9$ and the 8-spin unit cell
\cite{unitcell} used in our simulations is shown in Fig.
\ref{fig:lattice}. It can be viewed as consisting of loosely
connected $4$-spin plaquettes. Two topologically different types of
bonds can be distinguished. One are bonds within a plaquette $J_0$, the
other are dimer-bonds connecting plaquettes $J_1$. Additionally ${\rm
CaV}_4{\rm O}_9$ is believed to have a significant frustrating next
nearest neighbor (n.n.n.) antiferromagnetic interaction
\cite{discussion}.

Ueda et al. \cite{ueda} and Imada and Katoh \cite{katoh} have argued
that the spin gap can be explained as originating in a plaquette RVB
state, consisting of local singlets of the four spins on a plaquette.
This plaquette RVB state is the exact ground state in the limit $J_1=0$.
Second order perturbation theory around this limit suggests that it
survives even at isotropic coupling \cite{ueda,katoh}. A perturbation
around the dimer limit $J_0=0$ \cite{ueda} also leads to a wide range of
stability of the dimer singlet phase, but a small range of the couplings
exists, where no gap is observed in perturbation theory. First Quantum
Monte Carlo (QMC) results by Katoh and Imada\cite{katoh} also suggest
the existence of a finite gap $\Delta=0.11\pm0.03$ at isotropic
coupling.

Linear spin wave theory (LSW) \cite{ueda} and Schwinger boson mean field
theory results \cite{albrecht} on the other hand indicate that LRO could
survive at isotropic coupling despite the depletion of the lattice.
Exact diagonalization results are also contradictory
\cite{albrecht,sano}. They suffer greatly from the restriction to very
small clusters and the extrapolation to the infinite system size is
difficult. Sano and Takano\cite{sano} and Albrecht and Mila
\cite{albrecht} find a small spin gap, but also a substantial staggered
magnetization \cite{albrecht}. No definite conclusions can thus be drawn
from these calculations either.

To resolve these conflicting results we have determined the phase diagram
(see Fig. \ref{fig:phase}) of the non-frustrated model using the QMC
loop algorithm\cite{evertz}. Using this highly efficient cluster method
we can investigate larger systems at lower temperatures and with a much
higher accuracy than possible with the standard world line algorithm
used by Katoh and Imada \cite{katoh}. We have investigated lattices with
up to $N=800$ spins at temperatures down to $T=0.02$. The QMC method
suffers from no systematic errors and the results are reliable within
the statistical errors.

We find a wide region of stability of the Neel-ordered phase as a
function of the ratio of the couplings $\alpha=J_0/J_1$. We estimate the
lower boundary to lie between $0.55 < \alpha_c^l < 0.65$ and the upper
boundary between $1.05 < \alpha_c^u < 1.1$. At isotropic coupling LRO
thus survives the depletion of the lattice. The critical point
$\alpha_c^u$ is quite close to isotropic coupling, and a small
frustration might be sufficient to drive the system into the disordered
state.

To determine this phase diagram we have calculated both the spin gap and
the staggered magnetization. The spin gap $\Delta$ can be obtained from
the low temperature behavior of the uniform susceptibility $\chi$. Figure
\ref{fig:susc} shows $\chi(T)$ for some representative points. In a
gapped system it decreases exponentially as $e^{-\Delta/T}$ for low
temperatures.

Any finite system exhibits a gap, and thus a careful treatment of finite
size effects is necessary. For each temperature we have done
calculations on clusters of different size (up to $N=800$ spins) to see
whether our results have converged to the infinite system size limit. In
the regions of a large gap the convergence is quite rapid and it is no
problem to obtain the gap $\Delta$ from a fit of the low temperature
behavior of the uniform susceptibility $\chi$ to an exponential decay
$e^{-\Delta/T}$. In case of a vanishing or very small gap on the other
hand the susceptibility decreases linearly down to the lowest
temperatures $T_0$ we could study reliably on our finite clusters. In
these cases we cannot definitely decide about the existence of a gap,
but can only give an upper bound $\Delta<T_0$.

In Fig. \ref{fig:phase} we plot the gap obtained in this way together
with the perturbation theory estimates\cite{ueda,katoh}. Perturbation
theory is surprisingly accurate, but overestimates the gap slightly.
Specifically at the isotropic point we do not see any indication for a
gap, in contradiction to Katoh and Imada\cite{katoh}. Their calculation
of $\chi(T)$ is for a much smaller lattice (80 spins), and their gap may
be due to finite size effects \cite{katohqmc}.

The existence of LRO can be checked by calculating the staggered
magnetization $m_s$:
\begin{equation}
m_s^2 = \big\langle\psi\big|\left[{{1\over N}\sum_{\bf r}{\bf S}_{\bf r}
(-1)^{||{\bf r}||}}\right]^2\big|\psi\big\rangle.
\end{equation}
$m_s$ vanishes in the infinite system size limit in case of purely short
range correlations, while it is finite for LRO. The finite size scaling
of $m_s$ is known and a reliable extrapolation possible \cite{review}:
\begin{equation} m_s(N) = m_s(\infty) + {\rm O} \left({1\over
\sqrt{N}}\right) \end{equation}

Figure \ref{fig:moment} shows the system size dependence of $m_s$.
Let us first discuss couplings in the spin gap
regime. There the finite cluster results can be extrapolated linearly in
$1/\sqrt{N}$ to zero moment in the infinite system [Fig.
\ref{fig:moment}(a)]. In the double logarithmic plot [Fig.
\ref{fig:moment}(b)] it can clearly be seen that the results for finite
clusters bend down and approach the linear decrease (slope $1$). The
results for couplings in the LRO phase on the other hand bend up and
reach a constant value asymptotically. At the critical coupling itself
we expect a power law with a critical exponent different from the
${1\over\sqrt{N}}$-behavior of the gapped phase. A rough estimate shows
an exponent of the order $0.5$, as expected from the mapping to the
non-linear $\sigma$ model \cite{CHN}. This exponent is
indicated as a dotted line. A more detailed investigation, to obtain a
reliable estimate for the exponent and a better estimate for the
critical coupling is currently under progress.

Although the system size dependence is asymptotically linear in
$N^{-1/2}$, our lattices are not yet large enough to be really in that
limit. To get an estimate for the quality of our extrapolations we
extrapolate both $m_s$ and $m_s^2$. In case of LRO both extrapolations
should be linear.  We observe that, as seen in Fig. \ref{fig:moment}(a)
the system size dependence is not perfectly linear, but still bends down
a little bit.  Thus we take the value obtained from this fit as an upper
bound. In a plot of $m_s^2$ on the other hand a slight upwards bend can
be observed and we take that extrapolation as a lower bound. Both
extrapolations agree well. In the phase diagram (Fig. \ref{fig:phase})
we show the average value, with the error bars indicating these upper
and lower bounds. We have tested this procedure for the square lattice,
where our result of $m_s=0.306(3)$ agrees perfectly with the most
accurately known value $m_s=0.3074(4)$ \cite{wiese}. Again close to the
critical points the moment is very small and a definite decision about a
nonzero magnetization difficult. The magnitude of the staggered
moment compares well with the results of linear spin wave theory (LSW)
(also shown in Fig. \ref{fig:phase}), but the range of stability of the
LRO phase is overestimated by the LSW.

The conclusions obtained from the estimation of the gap and the
staggered moment are perfectly consistent. Starting from the dimer limit
we see a decrease of the gap as $J_0$ is increased. At
$\alpha=J_0/J_1=0.55$ we can still find a finite gap, while at
$\alpha=0.65$ we observe a finite staggered magnetization and a zero or
small gap. Thus we conclude that at a critical coupling $0.55 <
\alpha_c^l < 0.65$ the dimer singlet phase becomes unstable and the
model exhibits LRO. The critical coupling is probably close to
$\alpha=0.6$. There we cannot definitely decide about the existence of a
gap or LRO from our finite cluster results. Starting from the plaquette
side the gap also decreases as we increase $J_1$, but the plaquette RVB
state is stable for a wider range of couplings than the dimer
state. This is quite natural, as each spin is connected to one dimer
bond, but two plaquette bonds. Perturbation theory predicts that the
isotropic point is still in the range of stability of the plaquette RVB
state, but our QMC simulations show that LRO sets in at $1.05 <
\alpha_c^u < 1.11$. At the isotropic point we observe a substantial
nonzero staggered magnetization $m_s=0.178(8)$.

Comparing our results to previous calculations we find that the region
of stability of LRO is larger than estimated by second order
perturbation theory\cite{ueda}, but smaller than estimated by linear
spin wave theory and Schwinger boson mean field
theory\cite{albrecht}. Our results also agree well with the exact
diagonalization estimates of the staggered magnetization\cite{albrecht},
while the extrapolation of the spin gap data by exact diagonalization is
unreliable\cite{discussioned}.

By varying the ratio of the couplings in the $1/5$-th depleted square
lattice we can study both the LRO phase and the disordered phase of a
two dimensional quantum antiferromagnet, without having to introduce
frustration or to break symmetries, as in the dimerized square lattice
model \cite{dimer}. This model is thus ideal to study the critical
behavior and to test the predictions made by Chakravarty, Halperin and
Nelson based on the $2+1$-dimensional non-linear $\sigma$ model
\cite{CHN}.

In comparison to experimental results on ${\rm CaV}_4{\rm O}_9$ we
conclude that the depletion of the square lattice alone is not
sufficient to destroy LRO in the Heisenberg antiferromagnet, but it is
very close to the critical point. An additional frustrating next nearest
neighbor coupling is needed to drive the system into the gapped
plaquette RVB phase. All estimates from perturbation theory
\cite{ueda} and exact diagonalization \cite{sano} agree that
the stability of this plaquette RVB phase and the gap are greatly
enhanced by a frustrating next nearest neighbor coupling. Thus we expect that
already a quite small frustration will destroy LRO and can explain the
substantial gap observed in ${\rm CaV}_4{\rm O}_9$.

We would like to thank B. Ammon, H.G. Evertz, M. Imada, N. Katoh,
U.-J. Wiese and M. Zhitomirsky for helpful discussions. The QMC
program was written in C++ using a parallelizing Monte Carlo library
developed by one of the authors\cite{alea}. The calculations were
performed on the Intel Paragon massively parallel computer of
ISSP. M.T. was supported by the Japanese Society for the Promotion of
Science JSPS. We acknowledge financial support from a Grant-in-Aid
from the Ministry of Education, Science and Culture of Japan.


\input psbox.tex
\begin{figure}
\caption{The lattice structure of the depleted Heisenberg lattice
describing ${\rm CaV}_4{\rm O}_9$. Indicated are the two different types
of bonds, plaquette-bonds $J_0$ and dimer bonds $J_1$.}
\label{fig:lattice}
\end{figure}
\begin{figure}
\caption{Phase diagram as a function of the
ratio $J_0/J_1$. (a) shows the whole range of couplings. The
leftmost point corresponds to the dimer limit $J_0=0$ and the rightmost
point to the plaquette limit $J_1=0$. (b) A detail of the phase diagram
around the isotropic point plotted as function of $J_0/J_1$. Circles
indicate our QMC results for the spin gap, normalized by $J_0+J_1$. In
the gapless region the error bar indicates an upper limit for the gap.
Diamonds show the staggered magnetization. The error bars indicate the
upper and lower bound, as described in more detail in the text. As
reference we have included the perturbation theory estimates for the
gap\protect{\cite{ueda}} and the linear spin wave theory (LSW)
estimates for the staggered moment.}
\label{fig:phase}
\end{figure}

\begin{figure}
\caption{Temperature dependence of the uniform susceptibility $\chi$ for
different ratios of the couplings $J_0/J_1$. For each temperature the
system size was taken large enough to see the value for the infinite
system. The lowest temperatures were calculated on a $N=512$ spin
lattice. As a reference we have included results for the square lattice
Heisenberg model. The temperature is in units of the larger of the
couplings $J_0$ and $J_1$.}
\label{fig:susc}
\end{figure}

\begin{figure}
\caption{System size dependence of the staggered magnetization $m_s$ for
different ratios of the couplings $J_0/J_1$. For each system size the
temperature was chosen low enough to see the ground state properties.
The largest systems contain $N=800$ spins. (a) $m_s$ plotted as a
function of $N^{-1/2} $. A linear extrapolation gives the bulk
value. (b) A double logarithmic plot clearly shows the existence of long
range order or the linear decrease with system size
respectively. Included as guides to the eye are two straight lines
corresponding to power law decays with powers $1$ and $0.5$.}
\label{fig:moment}
\end{figure}


\begin{references}
\bibitem{unitcell} The smallest possible unit cell contains 4 spins. For
technical reasons we have chosen an 8-spin unit cell in our simulations.
\bibitem{discussion} There are still some discussions going on about the
magnitude of the couplings. It has been argued (see e.g. Ref. \cite{sano}
that the next nearest neighbor coupling could even be larger than the
nearest neighbor coupling. This question is of no relevance to the validity
of our results here, since the lattice obtained by considering only the
n.n.n. coupling are just two non-interacting copies of the same lattice.
The inclusion of the n.n. coupling on the other hand would then lead to a
different frustrating coupling, but the main effect, frustration would
be the same.
\bibitem{katohqmc} Katoh and Imada\cite{katoh} have
also estimated the spin gap from the difference of the energies of the
$S^z=1$ and $S^z=0$ subspaces. While this method is correct in the
$T\rightarrow0$ limit, their temperature of $T\approx0.1$, which is of
the order of their gap estimate, seems to be not low enough to
accurately obtain the ground state properties.
\bibitem{discussioned} Care must be taken in the extrapolation, since the
shape of the cluster is different for each cluster size. In fact the
$16$ and $24$ site clusters studied by Sano and Takano
\cite{sano} consist of $2\times 2$ and $2\times 3$ plaquettes. This
extrapolation should thus rather be viewed as a one-dimensional one to
the $2\times \infty$ limit.
\end{references}
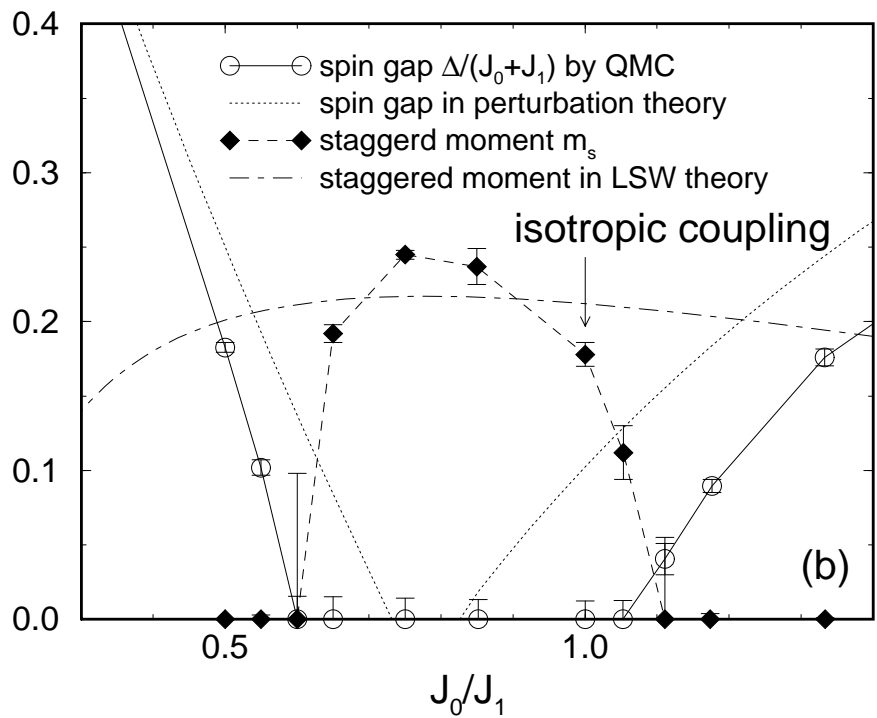
\end{document}